\documentclass{bmcart}

\usepackage[utf8]{inputenc} 

\usepackage{graphicx}

\startlocaldefs

\endlocaldefs

\begin{document}

\begin{frontmatter}

\begin{fmbox}
\dochead{Research}


\title{Answering open questions in biology using spatial genomics and structured methods}


\author[
   addressref={aff1},
   noteref={n1},                        
   email={sjena@fas.harvard.edu}   
]{\inits{SJ}\fnm{Siddhartha G} \snm{Jena}}
\author[
   noteref={n1},
   addressref={aff2},
   email={archit.verma@gladstone.ucsf.edu}
]{\inits{AV}\fnm{Archit} \snm{Verma}}
\author[
   addressref={aff2,aff3},
      corref={aff2,aff3},
   email={barbara.engelhardt@gladstone.ucsf.edu}
]{\inits{BEE}\fnm{Barbara E} \snm{Engelhardt}}


\address[id=aff1]{
  \orgname{Department of Stem Cell and Regenerative Biology, Harvard}, 
  \street{7 Divinity Ave},                     %
  \city{Cambridge},                              
  \cny{Massachusetts}                                    
}
\address[id=aff2]{%
  \orgname{Gladstone Institutes},
  \street{1650 Owens Avenue},
  \postcode{94158}
  \city{San Francisco},
  \cny{California}
}
\address[id=aff3]{%
  \orgname{Stanford University},
  \street{1265 Welch Road, x327},
  \postcode{94305}
  \city{Stanford},
  \cny{California}
}


\begin{artnotes}
\note[id=n1]{Equal contributor} 
\end{artnotes}

\end{fmbox}


\begin{abstractbox}

\begin{abstract}
Genomics methods have uncovered patterns in a range of biological systems, but obscure important aspects of cell behavior: the shape, relative locations of, movement of, and interactions between cells in space. Spatial technologies that collect genomic or epigenomic data while preserving spatial information have begun to overcome these limitations. These new data promise a deeper understanding of the factors that affect cellular behavior, and in particular the ability to directly test existing theories about cell state and variation in the context of morphology, location, motility, and signaling that could not be tested before. Rapid advancements in resolution, ease-of-use, and scale of spatial genomics technologies to address these questions also require an updated toolkit of statistical methods with which to interrogate these data. We present four open biological questions that can now be answered using spatial genomics data paired with methods for analysis. We outline spatial data modalities for each that may yield specific insight, discuss how conflicting theories may be tested by comparing the data to conceptual models of biological behavior, and highlight statistical and machine learning-based tools that may prove particularly helpful to recover biological insight.
\end{abstract}




\end{abstractbox}
%

\end{frontmatter}



\section*{Introduction}


The invention of the microscope allowed for unprecedented glimpses into the micron-scale world, and led to the first characterizations of the cell, including a view of the cell wall in plants and cell membrane in bacteria~\cite{mazzarello1999unifying}. A subsequent push to discover the constituent components of the cell led to the development of modern biochemical methods, predominantly based around density centrifugation. In this process, cells and tissues are dissociated and then separated by density to study the subcellular interactions between individual biopolymers. This approach progressively revealed the ``parts list'' of the cell, illuminating the composition of cellular structures such as the rough and smooth endoplasmic reticulum (E.R.)~\cite{fujiki1982isolation} and the Golgi body~\cite{ehrenreich1973golgi}. These methods, however, lost the spatial context of where biopolymers were located in the cell and the relative locations of the cells themselves. 

Imaging and biochemical characterization of cells and tissues have both made incredible progress since their initial development. Advances in physics in the second half of the 20th and early 21st century led to the invention of the electron microscope~\cite{koster2003electron}, scanning tunneling microscope~\cite{hansma1987scanning}, atomic force microscope~\cite{alonso2003feeling}, and super-resolution microscopy~\cite{schermelleh2019super}. Together with fluorescent proteins and affinity reagents (such as increasingly specific antibodies), these instruments opened a new frontier of molecular-level imaging. Scientists could interrogate the spatial location of different proteins, nucleic acids, or lipids within a tissue sample, and associate their distribution with particular cell morphologies or phenotypes. 

The development of high-throughput genomic sequencing technologies in the early 21st century led to the characterization of biology at base-pair resolution, first with bulk tissue samples as input, and later in single cells~\cite{mcgettigan2013transcriptomics, kulkarni2019beyond}. These protocols revealed the molecular composition of nucleic acids within tissues and cells, but without the spatial or visual context of imaging, since these methods required lysing cells to extract nucleic acids for sequencing.

The parallel technologies of sequencing and imaging have continued to increase in quality and resolution, and have complemented one another in important biological findings. A common post-hoc structure for leveraging the two approaches is to use statistical methods to identify correlations between an imaging-based readout and a sequencing-based readout~\cite{ash2021joint}, or predict gene expression levels in a sample using histology imaging~\cite{schmauch2020deep,comiter2023inference}; one such example is the mapping of somatic mutations, such as those found in cancer, to a cellular phenotype such as the emergence of dense cancerous tissue that is easily identifiable in pathology imaging~\cite{rios2017somatic}. More recently, pairing the two measurements in the same sample has become possible as biochemical methods to study genomics have expanded into the spatial realm. Fluorescence \emph{in situ} hybridization (FISH) methods involve probes that directly bind to proteins, RNA, or DNA of interest, allowing them to be imaged while preserving the location of the biomolecule~\cite{levsky2003fluorescence}. Alternatively, cells from a particular region of a tissue section can be sequenced together and reassigned to the tissue image afterwards, providing a coarse-grained view of cell-based gene expression across the tissue. Cells may also be optically barcoded prior to sequencing assays to capture their relative location. Using these methods, the high-dimensional genomic state of a single cell can be measured, and the cell can subsequently be mapped back onto its position in its native context, whether in culture or embedded in a tissue~\cite{bouwman2022era}; both cellular state and cellular environment is explicit in these approaches. 

Spatial genomics has already been used in a number of contexts to characterize genome-wide changes associated with cellular differentiation, development, interventions, and the progression of diseases such as cancer~\cite{rodriques2019slide, payne2021situ, zhao2022spatial}. With this new genome-scale spatially-resolved readout, researchers have the opportunity to discover general principles that govern cellular behavior in their environmental contexts. As better experimental methods are developed, of equal importance are the analytic frameworks that we use to understand and interpret the resulting spatial data. 

In this review, we take a look at the types of questions to which scientists may apply spatial methods, with an eye towards the methods appropriate for analyzing experimental results in the context of open questions in cellular biology. We first summarize the different spatial scales of analysis: molecular-, cellular-, and tissue-level data resolutions. We then examine four open questions that may be answered using spatial genomics 

\begin{enumerate}
    \item What is the functional spatial effect size of a cell?
    \item How does cell state and expression profile interact with cellular morphology, movement, and behavior?
    \item What local effects shape clonal dynamics in dividing and differentiating tissue?
    \item How does a cellular environment shape rare events?
\end{enumerate}
We review current and potential methods for answering these questions from spatially-resolved genomic data. As the suite of spatial genomics tools expands, we hope that the approaches discussed here may be generalized to a broad collection of robust, usable tools and data resources. 

\section*{Spatial scales of biology}
\begin{figure}[h!]
  \includegraphics[width=0.95\textwidth]{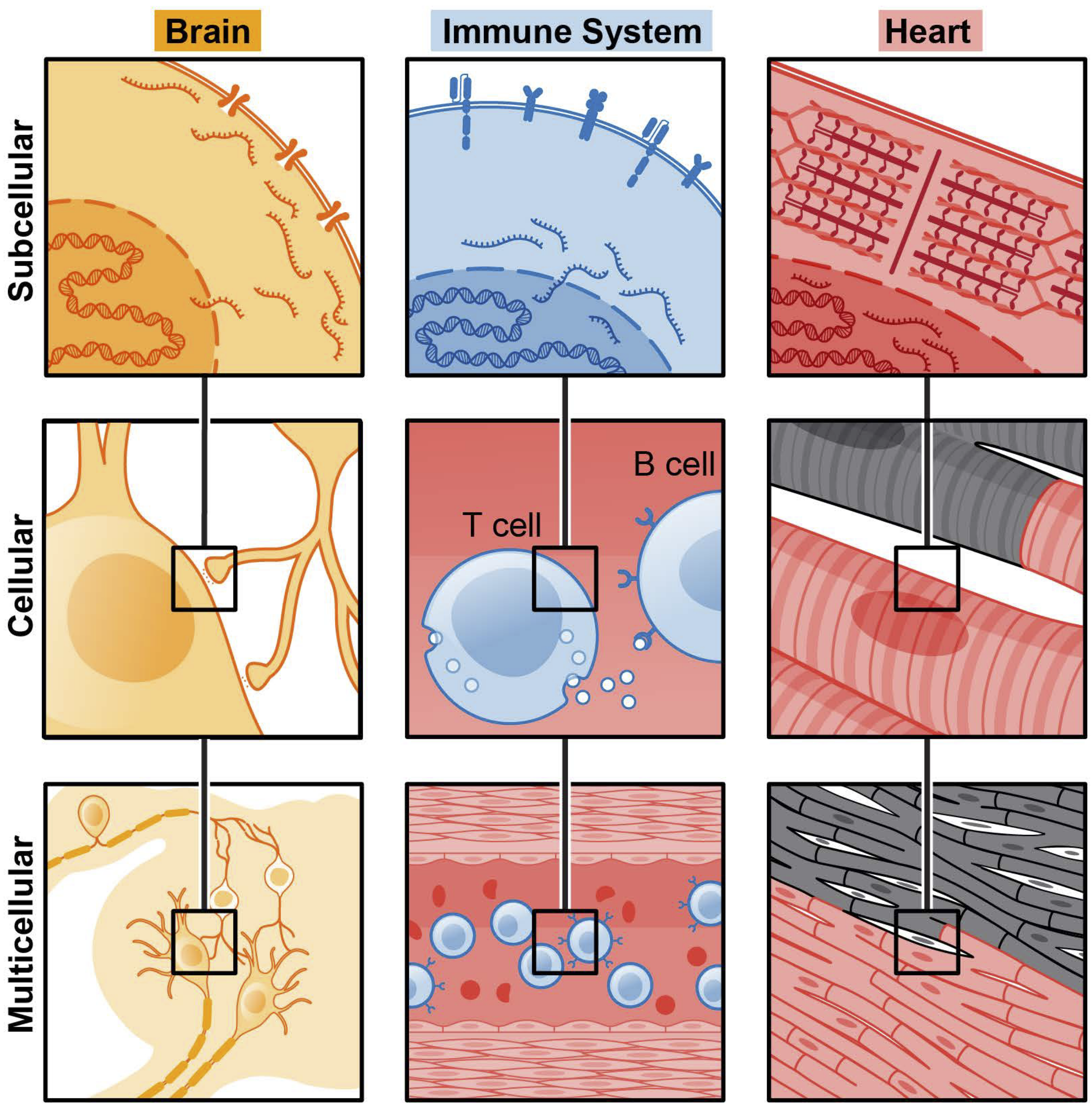}
  \caption{\csentence{Distinct scales of organization at different parts of the body.}
       Subcellular localization of
Receptors, cytosolic proteins, and signaling molecules affects cellular communication between neurons, B and T cells, or cardiac muscle in the heart. Each of these cell types is, further, a components of multicellular assemblies of many neurons, immune cells in the bloodstream, or heart tissue.}
      \label{fig:scales}
      \end{figure}

Peer through a microscope at a slice of tissue on a slide, and a wide range of cell shapes, sizes, and patterns present themselves. Further antibody staining reveals the location of proteins in specific intracellular compartments and throughout the extracellular matrix~\cite{thul2017subcellular}. A tissue sample contains biological processes occurring at three scales: sub-cellular (processes taking place within a subcompartment of a single-cell), cellular (processes taking place within $1-~10$ cells), and multicellular (processes taking place among $>10$ cells; Figure~\ref{fig:scales}). At the sub-cellular scale, our questions primarily involve interactions between individual molecules in organelles or membranes. At the cellular level, we ask questions about the overall composition of the cell and interactions with nearby cells. Finally, at the multicellular level, we ask how groups of cells of different types come together to form tissues with multifaceted functions.

\section*{Figures}

\subsection*{I. Sub-cellular resolution}
What molecules are in an individual cell and where do they function?  Nucleic acids and individual proteins are largely the drivers of cellular morphology and behavior. Using specific affinity reagents, such as antibodies or oligonucleotide probes, one can identify specific RNA and protein species in a fixed sample, providing insight into function. These molecules are often complexed together; one such example is chromatin, which consists of DNA, histone proteins, and often associated RNAs~\cite{van2012chromatin}. Here, we will describe some of the promising use cases for investigating these molecules at sub-cellular resolution.

\subsubsection*{DNA: Accessibility and structure}
DNA acts as the biological blueprint for an organism. With the exception of somatic mutations~\cite{martincorena2015somatic}, cells across an organism largely share the same DNA, yet serve vastly different functions. This is made possible through epigenetic modifications, which control the genes that are transcribed or repressed in a cell~\cite{goldberg2007epigenetics}. Structural changes from epigenetic modifications such as DNA or histone methylation~\cite{jones2001role, das2004dna, meissner2005reduced} can lead to differentially accessible regions along the length of the genome. These exposed chromatin regions, which may be read out through methods such as ATAC-seq~\cite{buenrostro2015atac}, allow binding of regulatory molecules such as transcription factors and RNA polymerase, leading to transcription. Other modifications, such as histone acetylation, can lead to recruitment of specific transcription factors and result in gene expression~\cite{goldberg2007epigenetics}.

Although distinct chromatin modifications have been associated with transcriptional activation or repression, the spatial organization of the genome and its link to the expression of specific programs remains less clearly defined. The genome is spatially partitioned, in structures largely driven by these epigenetic modifications, into domains of active or inactive genes called A and B compartments, respectively~\cite{belmont2022nuclear}. Although chromatin conformation capture methods such as Hi-C are able to capture these compartments~\cite{lieberman2009comprehensive, fortin2015reconstructing}, the link between these compartments and their transcription products is still being uncovered on a spatial level within intact cells. What occurs on the border between A and B compartments? Are there features that further distinguish different genomic compartments? A deeper understanding of spatial genome organization and its effect on the transcriptome would provide answers to these questions, as well as potentially addressing epigenetic dysregulation, which has been implicated in aging, response to environmental exposures, and disease progression~\cite{jones2015dna, zampieri2015reconfiguration}. 

\subsubsection*{RNA: Diversity and function}


Given the (generally) shared DNA sequences across cells from a single organism, variation in RNA expression is a major driver of cellular variation. Different cell types and cell states show different patterns of RNA expression, but spatial confinement of RNAs to compartments in the nucleus or cytoplasm is difficult to capture through conventional RNA sequencing. This is of particular interest since the dynamic organization of mRNAs can produce differential protein gradients in a tissue, driving processes such as metabolic regulation \cite{arceo2023translation}, polarization during \emph{Drosophila melanogaster} development or synapse formation in neurons~\cite{holt2009subcellular, lecuyer2007global, little2011formation}. Beyond mRNAs, a number of noncoding RNAs (ncRNA), such as long noncoding RNAs (lncRNAs) and microRNAs (miRNAs), have been identified and found to have important regulatory functions~\cite{laurent2015landscape}. Understanding the relationship between ncRNA function and their localization in specific nuclear and cellular compartments, combined with absolute transcript levels, would provide a more complete characterization of the transcriptional state of single cells~\cite{mckellar2023spatial}. Outside of the cell, RNA in extracellular vesicles may be implicated in inter-cellular signaling~\cite{mateescu2022phase}. Spatial transcriptomics provides insights previously unavailable that will further scientists' understanding of these RNA molecules.

\subsubsection*{Proteins: localization, abundance, and modifications} 
When possible, protein measurements provide the most direct window into active cell function. While the prevailing view is that transcript levels correlate with protein levels, possible discrepancies may arise between the two~\cite{li2020discrepant,wang2008discrepancy}, which may necessitate direct measurement of protein levels depending on cellular context~\cite{gry2009correlations, liu2016dependency}. Inferring protein levels from transcript levels also ignores aspects of protein regulation such as localization or post-translational modifications that may activate, modify, or suppress protein function~\cite{krishna1993post}. Antibodies to common protein modifications have allowed scientists to visualize cell processes such as signaling, while more extensive spatial measurements will allow for mapping of specific versions of proteins to subcellular locations within individual cells. 

\subsection*{II. Cellular resolution} 


Complex life is dependent on the cooperation and communication between many diverse cell types. Cell types are often categorized based on their interactions with other cells and tissues: for instance, neurons transmit signals to one another to form the basis of cognition~\cite{fries2005mechanism}, T cells identify and kill foreign cells~\cite{shevach2000regulatory}, and cardiomyocytes coordinate signaling between themselves to drive pacemaker activity in the heart~\cite{woodcock2005cardiomyocytes}. Recent advances in high-throughput single-cell measurements allow us to survey this diversity of cell types and interactions from a transcriptional or protein expression perspective. 

Variation also exists within cell type, often inelegantly lumped into the vague term \emph{cell state}. For instance, although a population of T cells are likely more similar to one another than they are to a muscle cell or skin cell, individual cells or subsets within the T cell population may be in different states of proliferation, activation, or quiescence at any given moment~\cite{shevach2000regulatory}. Cell states are controlled in part by local interactions between T cells and their environment, causing their transcriptomes and functional responses to diverge~\cite{li2020discrepant, dominguez2022cross}. Even in the absence of different environments, there are many subtypes of T cells each with their own cell state profile, and moreover cell states possess a natural level of variation within a population~\cite{verhoeven2022immune, dominguez2022cross}. Some of this variation is due to the stochastic nature of reactions such as transcription or chromatin dynamics occurring in single-cells~\cite{raj2008nature}. However, it is an open question how much of heterogeneity is random and how much is  byproduct of factors that are not measured in transcriptomic studies, such as interactions at the spatial boundaries of the cell population~\cite{snijder2011origins, raj2008nature}.   


Cells function together, so questions at the cellular scale must consider the interactions with individual cells in a local neighborhood. Receptor-ligand interactions that activate biochemical signaling pathways allow cells to modulate the transcriptomic state of nearby cells~\cite{almet2021landscape}. Cells from the same organisms may work together to perform complementary functions, like Schwann cells coating astrocyte neurons with myelin sheets to improve cell-to-cell signaling~\cite{nave2014myelination}. Cells from different organisms may also compete in the same tissue; for instance, immune cells fighting an infection or autoimmune diseases~\cite{baker2020emerging}.

\subsection*{III. Multicellular resolution}
In multicellular organisms, groups of many cells come together to form cellular complexes, tissues, and organs. Repeated patterns of multiple cell types in close proximity in a tissue are referred to as a \emph{niche}~\cite{liu2022spatial,marrahi2022four,medaglia2017spatial,tikhonova2020cell}. Beyond the identification and cell type composition of a particular niche, there is considerable interest in understanding niche sizing, the variation in niche architecture, the developmental trajectory of each niche type, and the interactions between niches~\cite{mayer2023tissue, pelka2021spatially}. For instance, stem cell niches are of particular interest as they possess the potential to regrow and regenerate specific tissues~\cite{lane2014modulating}.

Collections of niches create tissue architectures, and spatial transcriptomics presents the opportunity to bring more context to the organization of tissues from a molecular lens. How a repeated niche differs across the tissue may be explained by chromatin modifications or differences in RNA and protein expression, which lead to cell-type heterogeneity. Differences in structure from genetic defects can be explained causally by linking mutations to specific changes that propagate across the tissue~\cite{tepass1990crumbs}. 

The ambitions of single-cell studies have grown from defining distinct cell types~\cite{regev2018human} to the creation of comprehensive atlases -- from the tissue level~\cite{shahan2022single, verhoeven2022immune} to organs~\cite{sikkema2023integrated, nguyen2018profiling} to full organisms~\cite{tabula2018single, hung2020cell, wormatlas}, across age~\cite{taylor2019pediatric, chen2022single} and disease status~\cite{wilk2020single, schiller2019human, zhou2022alzheimer, grubman2019single, winkler2022single}. These atlases have the potential to advance biological discovery, in particular because they may provide a more thorough description of the distinct cell states in a larger population. Future work projecting single-cell atlases to spatial scales will add more context to these cell states.

\section*{Key questions in spatial biology}
\subsection*{I. What is the functional spatial effect size of a cell?}

\begin{figure}[h!]
  \includegraphics[width=0.95\textwidth]{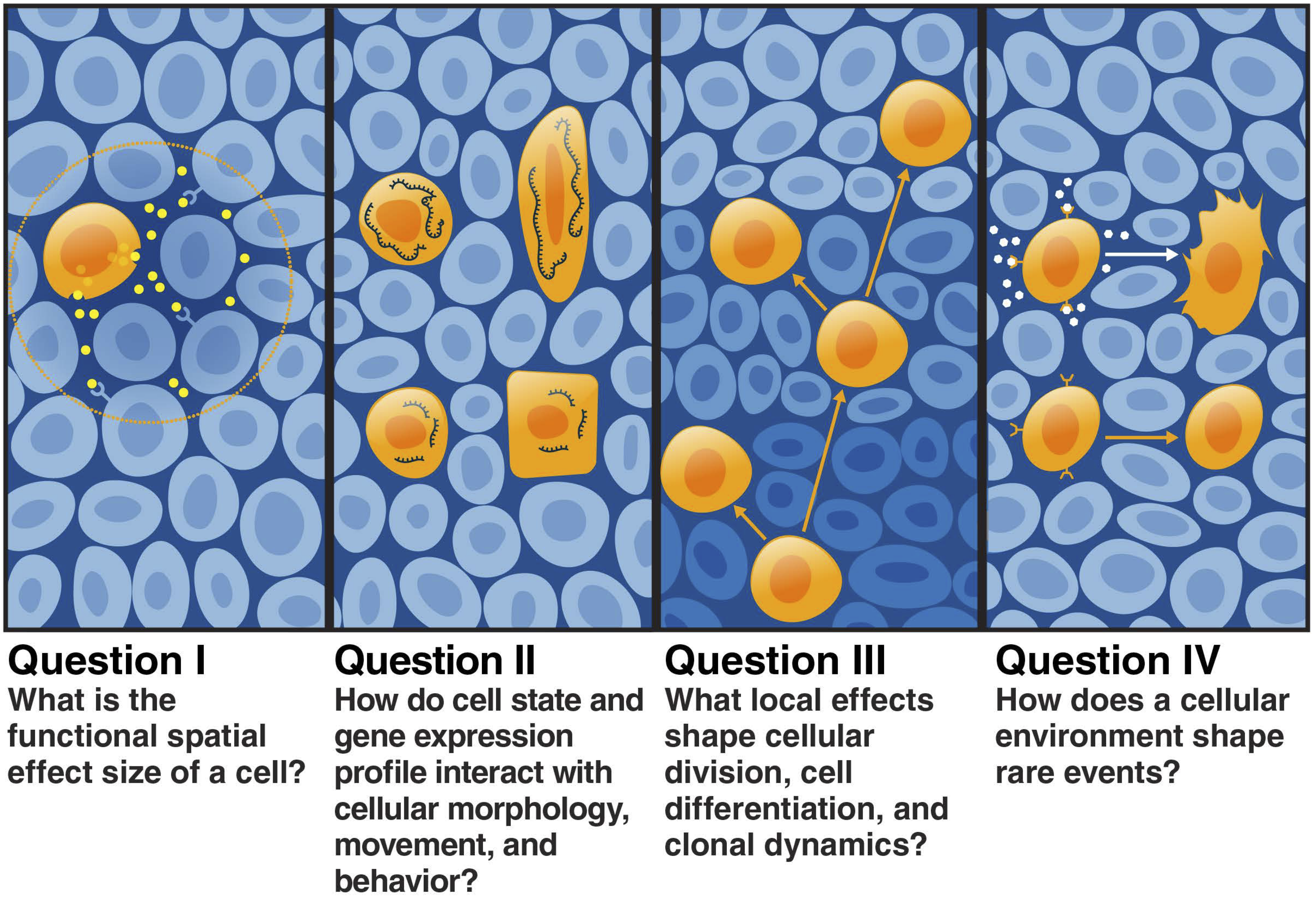}
  \caption{\csentence{Four key questions in spatial biology.}
     I. Cells can release ligands that allow them to communicate with
Other cells across various, unknown spatial scales. II. Cell location can affect morphology, movement of cells within a tissue and gene expression in unknown ways. III. It remains unclear how dividing clonal cells distribute within a tissue, and how this spatial distribution affects dynamics of gene expression. IV. It remains unclear how rare events in gene expression are influenced and orchestrated in within a tissue.}
      \label{fig:questions}
      \end{figure}

In a multicellular context, cells use many modes of communication to convey messages to their surroundings (Figure~\ref{fig:questions}, Question I). The mechanisms by which and extent to which cells are able to communicate across the body has long captivated biologists. The concept of morphogens or hormones to explain cells communicating over distances is an old one~\cite{starling1914discussion}. As biochemical tools grew more sophisticated, numerous signaling molecules and pathways were found to serve this critical role~\cite{nair2019conceptual,levi1951selective,hokin1953enzyme,krebs1956phosphorylase}. These signaling pathways, which are often conserved across organisms, continue to inform research today; for example, live-cell imaging revealed that the Ras/ERK pathway propagates waves of signaling activity during development in response to processes within the cell as well as directing events such as wound healing that take place outside of the cell~\cite{aoki2013stochastic, mcfann2022putting, marmion2023stochastic}. As we piece together the toolbox of molecules used for cellular communication, it remains unclear how to best measure coupling between cells in a tissue. For a given cell, how do we know how much of its behavior is due to communication with cells around it? How far does this communication stretch?

Biophysical models of cell interactions form a useful framework within which to ask these questions. Perhaps the earliest of these models is the the French Flag Model of morphogen gradients, where particular levels of a molecule are mapped to distinct outcomes in a tissue. This provided a conceptual explanation of how diffusing gradients of such a chemical could result in patterning along the length of a developing embryo. This class of model led to more mathematically sophisticated descriptions such as the Turing Model, which describes pairwise interactions between two molecules that are capable of generating stable patterns in a tissue~\cite{green2015positional}. In these early models, the parameters of interest were i) the geometry of the tissue, ii) the level of morphogen(s) in space and time, iii) the feedback, feed-forward, or cooperative interactions occurring between morphogens in space, and iv) the readout of cell state as a function of these morphogens. As more interesting biological patterning questions emerged, modeling these behaviors expanded accordingly. Ising models---spin models based on a lattice-structured Markov random field adapted from particle physics--- that have similarly been used to think about cellular interactions~\cite{merle2019turing}. Kuramoto oscillators model coupled cells with continuous states to drive phase differentiation~\cite{breakspear2010generative}. Alternatively, information-theoretic approaches been used to understand how small sets of signaling genes can encode a rich space of spatial architectures from experimental data, combining mathematical biophysical models with experimentally collected data~\cite{petkova2019optimal, tkavcik2008information}. However, throughout all of these modeling paradigms, the key components of cell connectivity, morphogen levels, and morphogen interactions have remained unchanged.

Spatial sequencing provides a high-dimensional dataset to statistically identify the genes involved in intercellular communication in different contexts. Early analysis has focused on identifying and building on known ligand-receptor pairs. In the analysis of the initial seqFISH+ results~\cite{eng2019transcriptome}, the authors looked for enrichment of expression of known ligand-pairs pairs in adjacent cells by comparing null expression distribution created by permutation shuffling. On the same data, graph convolutional neural networks were trained to predict the probability of two genes interacting given the spatial neighbors graph and expression of the two genes in each cell~\cite{yuan2020gcng}; known ligand-receptor pairs were used as positive and negative examples. Optimal transport methods were used to identify similar distributions of known receptor and ligand expression patterns in spatial data~\cite{cang2020inferring}, which captured potential interactions beyond spatially-adjacent cells. 

Although limiting analysis to known receptors limits the investigator's ability to discover new signaling motifs, testing pairs or higher order sets of genes for interactions leads to combinatorial increases in statistical tests and computational demands. Few experiments have sufficient sample size to adequately power investigations of high order interactions. Instead, statistical methods often jointly model all genes together and try to learn larger groups, or ``topics,'' of co-varying genes that are associated with spatial patterns. For example, Gaussian process regression can model spatial gene expression with clever kernel composition~\cite{arnol2019modeling}. Three kernels are used to decompose gene expression variance into intrinsic cell effects, extrinsic or environmental effects, and cell-cell interactions. Comparison to a null model assuming no cell-cell interactions identifies communication-related genes.  Related work reconstructs gene expression from given cell-type labels and spatial neighbor graphs using autoencoder architectures~\cite{fischer2022modeling}. While not explicitly using the expression levels of other genes, the cell-type label serves a similar role in capturing expected nonspatial gene-gene correlations. This work similarly uses a null model without spatial connectivity to test for interacting genes. For both strategies of testing pairs of genes or a gene against all other genes, the number of tests done requires proper null models, multiple hypothesis testing correction, and a reliable way to control for known cell-type heterogeneity and adjacency, which bias the results. 

Nevertheless, single snapshots of spatial expression data sets miss important information on the temporal nature of signaling. Parameters such as the responsivity of a cell type to a particular protein, or the pairwise interactions between two genes, may change as a function of time. For example, spatial measurements at two stages of uterine development were used alongside CellPhoneDB~\cite{vento2018single, efremova2020cellphonedb}, a database of know ligand pairs, to identify which cell types had compatible signaling proteins and were likely to be in communication over development~\cite{garcia2021mapping}. Increasing the resolution of time points will allow the expansion of such statistical techniques to identify interactions over time. Time-series analyses can also help identify more causal relationships in signaling. Fluorescent live-cell imaging data and point-process models were used to quantify ERK signaling and downstream Fos expression in different drug conditions~\cite{verma2021self}. Specifically, self-exciting Hawkes processes model expression and signaling among cells over time and space. Newer fluorescence imaging protocols will expand the number of behaviors that can be captured simultaneously in live-cell imaging data~\cite{borjini2019imaging}.

Understandably, we currently have the most confidence in interactions between directly adjacent cells, since long-distance channels or indirect secretory pathways through which cells can send or receive messages are more difficult to study. By incorporating multiomic measurements of cells in space, as well as potentially integrating time-resolved measurements, we may be able to better understand cell communication at a distance.

\subsection*{II. How do gene expression profiles interact with cellular morphology?}

A common practice in both basic cell biology and pathology is to use cell morphology to distinguish cell types or states. Cellular structure informs function, and thus cells from different tissues and different cell types in a single tissue vary markedly in their appearance (Figure~\ref{fig:questions}, Question II). For instance, due to disregulation in growth pathways, cancerous cells are commonly larger than their healthy counterparts, and are often more motile when imaged over time under a microscope. Physical stress can also change cell state and gene expression; mechanical stretching of fibroblasts was found to dramatically alter epigenome states to enable cells to prevent damage to the physical structure of the genome~\cite{nava2020heterochromatin}. 

Before spatial single-cell technologies, some methods attempted to jointly model morphology patterns and gene expression from paired bulk tissue samples~\cite{ash2021joint, gtex2017genetic}.  The clear limitation here is the mismatch between resolutions: Images can provide cell-level phenotypes but bulk expression cannot. The emergence of spatial single-cell measurement techniques is poised to overcome this limitation.

The primary open question here is how to best represent morphology. While gene expression is conventionally represented by a count matrix, there is not a similar universal tabular form to represent morphology. Recent approaches have attempted to provide solutions to this problem. One strategy is to convert morphological data, generally in the form of images, into tabular data of derived features. One study measured gene expression with the L1000 assay and morphological features such as nuclear area or DNA organization using the Cell Painting assay~\cite{haghighi2022high}. Lasso logistic regression and a multi-layer perceptron accurately predict gene expression from $\sim$1000 morphological traits provided by CellProfiler. CellProfiler was also used to create tabular readouts from paired imaging and single-cell CRISPR-Cas9 knockouts, in order to cluster gene knockouts with similar morphological changes and build genetic interaction networks~\cite{ramezani2023genome}.

An alternative approach is to use neural networks to capture the features of an image. For example, MorphNet uses a convolutional general adversarial network (GAN), which uses two neural nets in competition to improve performance, in order to predict morphology from gene expression in brain-wide MERFISH data~\cite{lee2022morphnet}. A recent study performed a similar paired CRISPR knockout and imaging dataset to Ramezani et al. (2023) but calculated embeddings of images from a self-supervised vision transformer trained on single-cell Cell Painting images~\cite{sivanandan2023pooled}. This approach outperformed a classic image featurization representation for classifying single CRISPR perturbations' mechanism of action and recovering known biological relationships between genes~\cite{sivanandan2023pooled}. 

Each approach has different strengths and limitations. Individual features derived from cell painting methods are easier to interpret and can be used in small sample size regimes. Tabular data is amenable to traditional statistical regression methods and the accompanying theoretical guarantees; however, count data and specific experimental designs often require additional structure on the methods that are challenging for non-statisticians. Neural networks provide more flexibility in the morphological variation that they capture, but require both an adequate amount of data for training and some expertise in training and interpreting deep learning models. Both approaches also require identifying which variable should be the output and which variable should be the input. 

Biologically, cell morphology and movement are determined largely by membrane contacts; cell membranes are predominantly composed of lipids and proteins, and the abundances of these components are largely dictated by gene expression. However, changes in morphology and motion also drive changes in gene expression as the cell responds to new conditions. Many signaling pathways begin with external influences on membrane proteins. These feedback loops suggest the causal relationship goes both ways, limiting static data to providing mostly correlative findings.

In the near future, decreases in costs and improvements in resolution will allow scientists to better establish the causal relationships between gene expression and morphology. Time-series measurements and live-cell imaging can uncover the temporal ordering between gene or protein expression events and morphological changes. Single molecule tracking will be able to resolve where in the cell proteins are functioning and creating structural features~\cite{mcswiggen2023high}. These improvements will also shine further light on the relationship between morphology, motion, and function. With improved experimental methods and proper statistical techniques, a complete understanding of the determinants of cell morphology seems within grasp.

\subsection*{III. What local effects shape clonal dynamics of dividing and differentiating cells?}

Cell division establishes populations of clones in various tissues around the body (Figure~\ref{fig:questions}, Question III). Division can maintain genomic, transcriptomic, and epigenomic information, but comes with the downside of passing on potentially deleterious properties such as DNA mutations and aberrant epigenomic states. On the other hand, precise maintenance and expansion of particular clones underlies important processes such as the development of adaptive immunity. Some biological processes are ``bottlenecked'' in the sense that unfit clones die out due to physiological conditions~\cite{sun2014clonal}. However, many cell populations, including hematopoetic stem cells that give rise to the entire lineage of circulating blood cells, are comprised of hundreds or thousands of clonal populations, including clones that harbor mutations that decrease proliferation~\cite{fabre2022longitudinal}, suggesting that clonal heterogeneity may be the rule rather than the exception.

Biophysical models of clonal dynamics have been studied for many years in the context of stem cells. A primary question for any stem cell population is whether and how the population replenishes. This has been modeled by three parameters representing three distinct probabilities of division outcomes for a stem cell: 1) division into two stem cells, 2) two differentiated cells, or 3) one stem cell and one differentiated cell~\cite{klein2007kinetics, klein2008mechanism}. Recently, another intriguing possibility has been introduced: rather than remaining in static states, cells may stochastically transition between stemlike and differentiated states with some probability, only fully converting to a distinct fate when faced with a particular signal~\cite{parigini2020universality}. 

These relatively simple ``state transition'' models, applied to well-mixed or spatially structured populations, have been used to great effect to predict behavior of stem cell populations. Crucially, certain regimes representing distinct probabilities of differentiation or division can be distinguished from one another experimentally through the resulting predicted distributions of clone sizes. One early method for marking clones involves dosing subsets of cells with a dye that subsequently becomes diluted over time, a method that is commonly used to monitor T cell proliferation~\cite{gudmundsdottir1999dynamics}. While this can accurately mark the generation, it does not provide direct linkage across generations. 
Another method involves inducing fluorescent protein expression in a subset of cells, using this to identify groups of fluorescent cells that all originated from the same clone. Similarly, this approach does not allow for identification of mother-daughter cell relationships, but can be used to measure clone size distributions by quantifying the size of distinct groups. 

Experimental methods to identify mother-daughter relationships between single cells within a clonal population, on the other hand, have been difficult until CRISPR/Cas9 was developed. DNA-based barcodes for clonal tracking are an attractive technological development towards addressing clone-related questions. Static barcodes can be introduced into the genomes of cells in a random fashion, so that some distribution of barcodes is introduced into the first generation and subsequently passed on to each cell's progeny~\cite{gerrits2010cellular, nguyen2015barcoding}. Through subsequent DNA sequencing, the barcode for each cell can be recovered to establish clusters of cells that arise from the same clone. More recently, dynamic barcoding can be used to establish precise lineages: in this method, CRISPR/Cas9 randomly edits a barcode as it is passed on from cell to cell, allowing researchers to reconstruct lineages through the introduction of random SNPs~\cite{spanjaard2018simultaneous}.

\begin{figure}[h!]
  \includegraphics[width=0.85\textwidth]{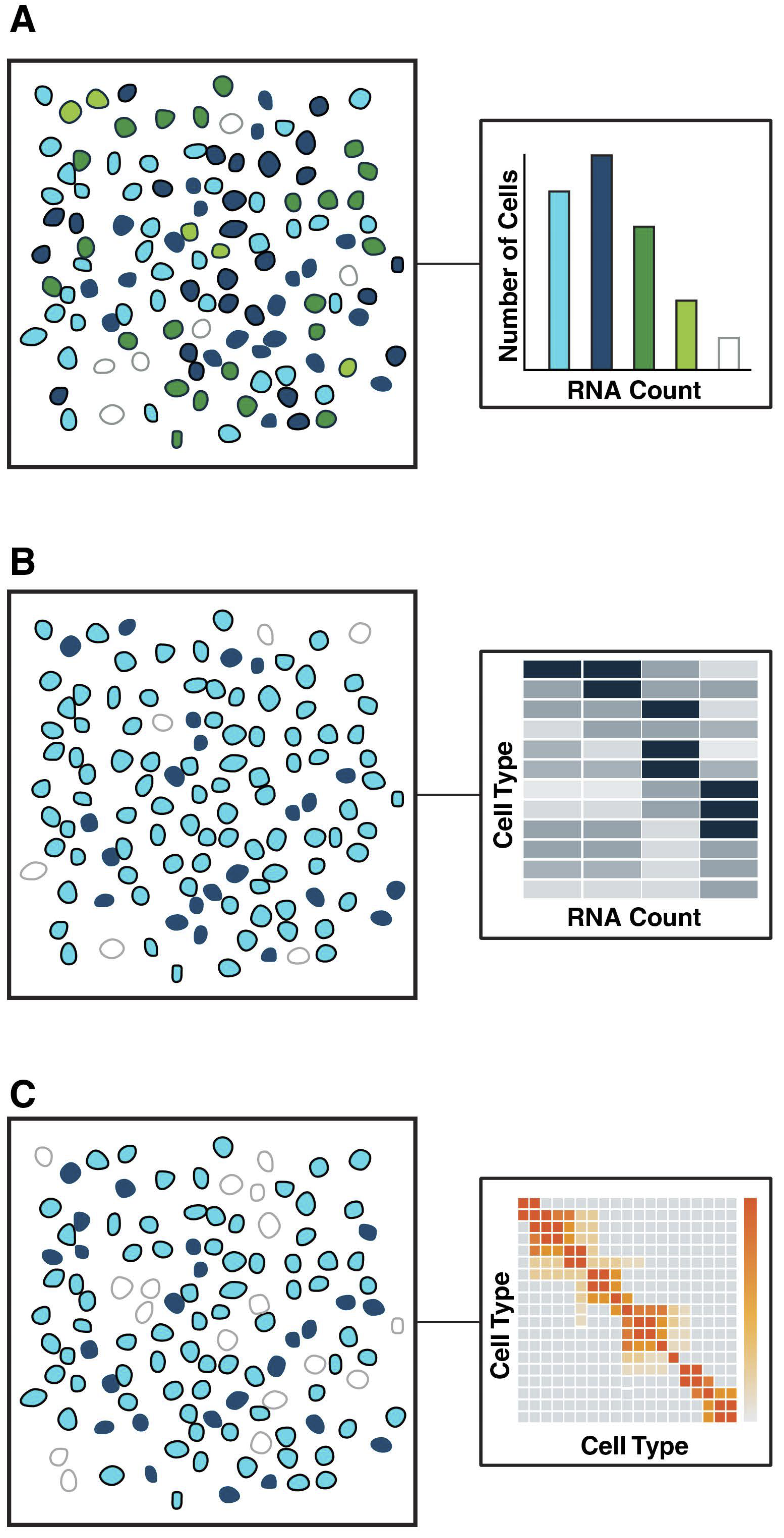}
  \caption{\csentence{Essential cellular behaviors assayed in spatial genomics.}
   Distributions of RNA (A), cell type clustering from gene expression (B) and spatial correlations (C) can all be measured from spatially resolved sequencing data.}
      \label{fig:behaviors}
      \end{figure}
      \clearpage

Combining imaging-based methods with barcoding offers an opportunity to build models of clonal expansion in a spatial context. In particular, work extending clonal dynamics models to the mammalian epidermis exposed complex emergent clonal behavior that arises when cells are confined to a tissue~\cite{klein2007kinetics, klein2008mechanism}. The epidermis is highly stratified, and, within a layer, clonal populations of stem cells can often be visualized as spatially defined clusters of mitotically active or inactive cells. Post-mitosis, differentiated cells that arise from a stem cell on a basal layer will often emerge on a suprabasal layer, giving rise to complex geometries of clone dispersion spanning three dimensions~\cite{koster2007mechanisms, colom2020spatial, colom2021mutant, blanpain2014plasticity}. Specific functional geometries of tissues, such as the crypts of the stomach or the cylindrical structure of vasculature, likely involve similarly unique geometric patterning of clones (Figure~\ref{fig:behaviors}).

We expect that interrogating clonal populations in their native tissue through a combination of imaging, barcoding, and transcriptomics will allow for a broader range of clonal behaviors to be defined. In particular, although clonal populations tend to be ``coarse-grained,'' as observed in the epidermis as well as in metastatic clones using spatial DNA sequencing~\cite{zhao2022spatial}, it remains to be seen how ``fluid'' individual clonal populations are within a tissue. 

Are there definable subclones within a clone that occupy their own spatial niche? In the case of cancers, cells from one clone may metastasize to form their own population elsewhere. In what ways is this subclone distinct from the original? Prior work used variance decomposition of Slide-seq data to identify gene signatures that explained differences between distinct clones as well as subclones within cancerous tissue~\cite{ma2022spatially}; similarly, constrained regression and covariance estimation were used to study clonal populations using copy number variation~\cite{ru2023estimation}.
Related work jointly identified copy number polymorphisms in spatial transcriptomics and inferred cellular clones in tissues using a hidden Markov random field~\cite{elyanow2021starch}.
Extending spatial experiments using dynamic barcoding would allow for fine-grained resolution of subclone emergence in the future; analytic methods to reconstruct the full clonal trajectories would add specific mother-daughter relationships in space.


\subsection*{IV. How does a cellular environment shape rare events?}
The first single-cell RNA-seq datasets confirmed what many biologists had already suspected: that substantial expression heterogeneity exists between cells in a tissue, and that this heterogeneity underlies a wide range of diseases. For instance, cancers often arise not as a function of cellular collectives, but as a function of one particular cell. A dominating paradigm in cancer is that single cells experience a ``perfect storm'' of factors that leads to them becoming \textit{jackpot cells}, or clones that express a specific mRNA at extremely high levels where their sister clones express none~\cite{shaffer2017rare} (Figure~\ref{fig:questions}, Question IV). In some cases, these rare cellular states are transient: jackpot cells may not always express combinations of genes throughout their lifetime, and may not pass on their phenotypes to their progeny. In BRAF melanoma, jackpot cells fail to follow Luria-Delbruck behavior and do not pass on their properties unless challenged with the addition of a drug, which then stably locks in the resistant state~\cite{shaffer2017rare}. This implies that every time a population of cancer cells is challenged with a drug, a constant but small fraction of the population experiences stochastic resistance. Other rare cellular phenotypes are more consistent with Luria-Delbruck dynamics; for instance, rare mutations causing cancerous growth are passed on from mother cell to daughter cell to create large colonies and eventually solid tumors~\cite{reiter2018minimal}. 


While we are beginning to understand the factors affecting jackpot cell emergence in culture, the environmental factors (e.g., tissue niche, surrounding cellular milieu, position in the tissue) that regulate the cell states giving rise to heterogeneous gene expression events are still unknown. Leveraging spatial genomics to identify these rare events such as jackpot cells among other cells in a tissue may lead to a better understanding of these rare events. However, a major limiting factor in studying such rare events is statistical confidence in detecting such events. In studies performed on melanoma cells, jackpot cells were detected using RNA-FISH with probes targeting a pool of pre-identified drug resistance genes~\cite{shaffer2017rare}; this allows for high confidence calls of jackpot cells that may not be possible in standard single-cell sequencing workflows. The total number of mRNA transcripts per cell is typically much lower than the mRNA counts collected using FISH methods, especially in such a small pool of target genes. In this particular case, bulk RNA-seq was used to identify a set of high-confidence genes for RNA-FISH probing. However, the candidate genes designating jackpot cells may not always be so well defined, and using a sparse readout such as single-cell transcriptomics to identify novel jackpot cells presents a circular problem.

\section*{Methods opportunities in spatial biology}

\begin{figure}[ht!]
  \includegraphics[width=0.95\textwidth]{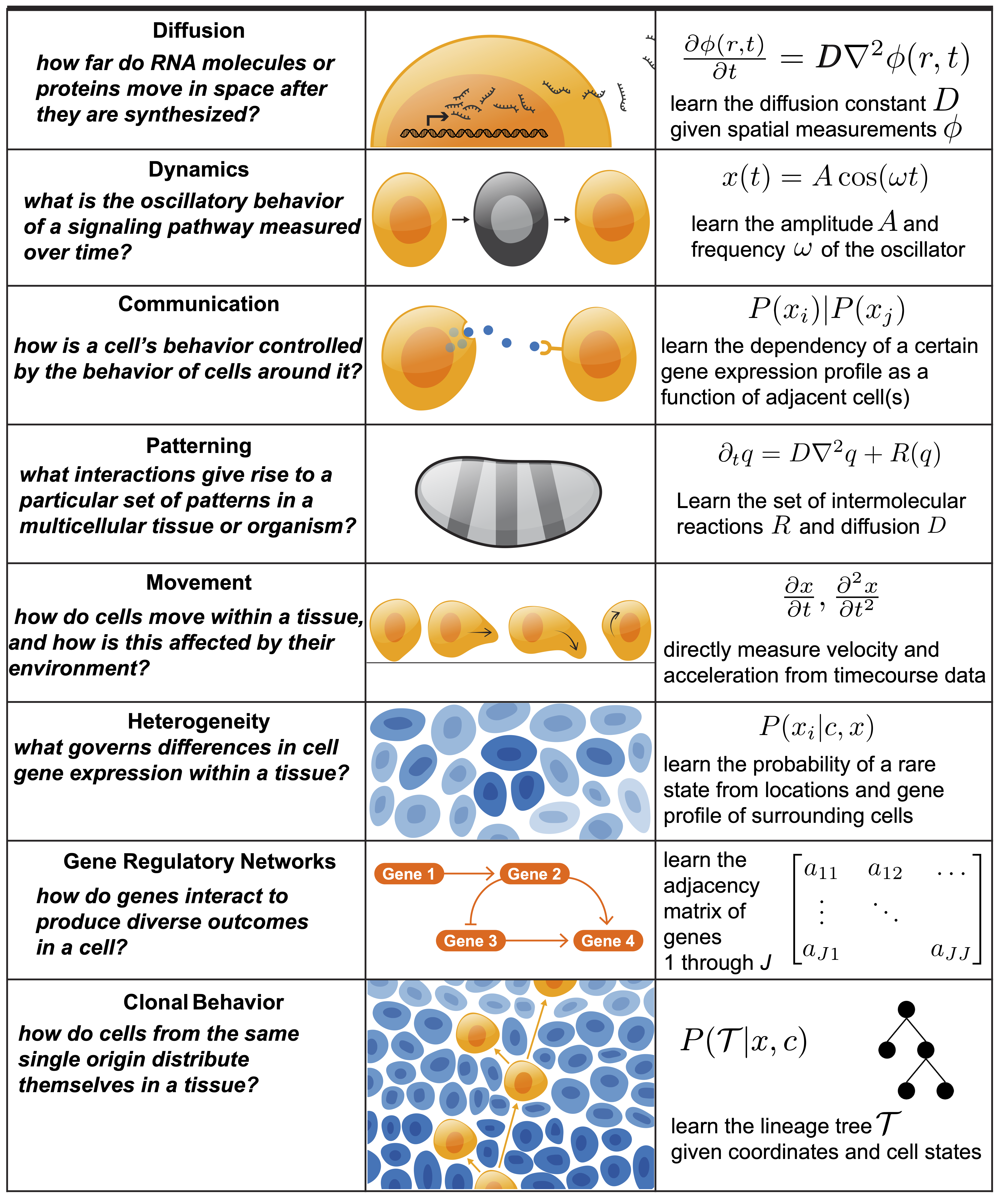}
  \caption{\csentence{Methodological opportunities for spatial genomics.}
    We describe distinct ``classes'' of biological and biophysical measurements that fall within our four key areas of interest. These include diffusion of RNA away from the site of transcription, establishment of patterning in a multicellular tissue or organism, and gene regulatory networks giving rise to particular behaviors. For each, we describe how the underlying processes may be directly measured, or indirectly inferred, from spatial genomics data. }
      \label{fig:opportunities}
      \end{figure}

These four open questions in spatial biology---along with the existing or forthcoming technologies to observe these phenomena in tissues---require the development of statistical approaches to arrive at satisfying answers. The opportunity here is in building models that incorporate additional structure -- time, space, or environmental context. Here, we outline opportunities for methods development in each of the four areas, focusing on methods that are most likely to be successful given the constraints of the data and sample size (Figure~\ref{fig:opportunities}).

To illustrate the structure of potential novel and existing methods, we assume that we start with one of two structured datasets. The first dataset is two tables, a cell by gene (or other feature) count matrix $X \in \mathcal{R}^{N \times G}$ and cell by spatial coordinate matrix $C \in \mathcal{R}^{N \times D}$, where $N$ is the total number of cells assayed, $G$ is the number of genes assayed, and $D$ is the number of spatial dimensions (this will almost always be 2 or 3). We will use $x_{i,j}$ to refer to the count of gene $j$ in cell $i$, $x_{i,-j}$ to represent the gene counts in cell $i$ of genes other than $j$, $x_i$ to refer to the vector of all gene counts in cell $i$, and  $x_{-i}$ to refer to all gene expression in cells other than $i$, with similar subscripts for the coordinates. 

Alternatively, we may have a more granular set of observations of the identity of each of $M$ molecules observed (e.g., spatially localized RNA transcripts), with a location for each molecule $c_m \in R^{D}$, and the cell it belongs to $o_m \in {1,2,\dots,N}$. We will use $c_i$ to loosely refer to the coordinates of all molecules in cell $i$ and $m_i$ for the identity of all molecules in a cell $i$. In the following section, question-specific data and notation will also be introduced to illuminate the modeling approaches proposed. For each opportunity, we try to identify challenges across data collection, model architectures, and model inference and evaluation.

\clearpage

\subsection*{I. Methods to characterize the functional spatial effect size of a cell}

A spatial experiment observes an instance from some distribution over the expression and spatial coordinates of the cells, $p(X, C)$. Signaling between cells implies there is some conditional relationship of a cell's state on other cell's state. A model to identify spatial signaling assumes that the variability of cell state can be decomposed into factors from other cell states (\emph{extrinsic factors}) and cell-specific factors (\emph{intrinsic factors})~\cite{elowitz2002stochastic,swain2002intrinsic}. This may look like a model with form:
$$p(x_{i,j}|x_{i,-j}, x_{-i}, C) = f_1(x_{i,-j}) + f_2(c_i) + f_3(x_{-i}, C).$$
 
We use $f_1$ to represent how cellular state feature $j$ is dependent on the other state features in the cell (i.e., intrinsic factors). Often, cell type is used as a proxy for intrinsic cell state. Some lower-dimensional embeddings, such factor analysis, fit without spatial information can also be used to find the intra-cellular covariance between features of cell state.

Next, $f_2$ represents a spatial pattern of cell state that is a function of location but not environment. This variability may reflect some global architecture of cell types and niche organization. It accounts for variation in cell state that is not part of the signaling pattern we are attempting to find. For example, a tissue sample might be organized with different cell types separated in distinct regions of space, which creates a spatial pattern of gene expression that is not the result of (short term) signaling behavior. We can think of a model like spatial nonnegative matrix factorization \cite{townes2023nonnegative} as decomposing the variance among these two terms: non-spatial factors capture the intra-cellular covariance while spatial factors learn the spatial archetypes for each feature of cell state. 

The final term, $f_3$, represents perhaps the most interesting behavior: the dependence of cell state on local cells. We are looking for repeated variation in cell state that cannot be explained by the other features in a cell or by global patterns of expression. Of critical importance is correctly teasing out this relationship from our spatial term $f_2$. This can be done by restricting the cell's dependence to only neighboring or nearby cells, making $f_3$ represent the ``local'' covariance of gene expression. 

As observed before, spatial factor models tend to capture only the first two functions while missing local signaling effects~\cite{velten2022identifying, townes2023nonnegative}. Looking at the correlation between known ligand-receptor pairs expression across neighbors uses cell type to control for the effects $f_1$ and proximity to zero out $f_2$, testing specifically for the existence of $f_3$. 

Thus the opportunity remains to fully model all three factors simultaneously. Data with distinct local and global signals will need to be used for a model to learn the desired patterns. The appropriate functional forms of each term will be required to accurately capture biological processes; nonlinear functions will likely be necessary for an accurate model, although they will increase the difficulty of inference and also the data requirements for adequate power. For $f_3$, given the most obvious adjacency heuristics, models can estimate signaling between adjacent cells, but more complex communication across larger spatial scales may be hard to detect effectively. Ideally, these models can look at signaling across all features, though computational complexity may require low-dimensional latent factor representations to tractably modeling complex signaling. Bayesian representations can provide proper uncertainty quantification and identify multiple parameter ``optima'' that explain the data equally well -- but also require more expensive computation of posterior distributions. 

Using the specific location of molecules, our second data representation allows for increased granularity and ability to look for causal signals. Here, models can explicitly condition on the location of a molecule inside or outside a cell as a proxy for determining its contribution to signaling behavior. Proteins near the membrane, for example, are more likely to be involved in some extracellular signaling than nuclear proteins. In these cases, the coordinate of a molecule might be considered rather than the cell center: 
$$ p(m_i, c_i|m_{-i}, c_{-i}, o_{i}, o_{-i}) = f_1(o_i) + f_2(c_{n/-i}, m_{n_/-i}) + f_3(m_{-n}, o_{-n}, c_{-n})$$

Here, $f_1$ is dependent on the cell type, positing some shared spatial organization across cells of the same type. $f_2$ is accounting for some variation from the organization of the other molecules in the cell and $f_3$ is accounting for variation from molecules outside the cell. In this setup, $f_2$ captures intra-cellular signaling, perhaps of some cascading pathway, and $f_3$ captures inter-cellular signaling.

Like the models based on cell count tables, opportunities exist to model local and global effects at molecular resolution. Data with both intra-cellular and inter-cellular behavior will be needed to calibrate the effectiveness of such a model, though the identification of known pathways can serves as a model evaluation metric. Similar computational challenges in terms of dimensionality of possible gene-gene interactions will plague these kinds of models, which may require the development latent variables models of single-cell spatial organization. 

For both approaches, we are missing an important variable in the time dependency of signaling. A spatial measurement only provides a snapshot of the present; some molecules may be moving towards their destinations while others with important interaction effects may have just degraded. The current location may not be entirely informative about the relevant signaling actors. As multiple spatial snapshots and live-cell imaging become more affordable and widespread, models that explicitly include dynamic behaviors will be invaluable for establishing causality in biological signaling processes. 


\subsection*{II. Methods to investigate the relationship between morphology and expression}

When biologists study the relationship between morphology and expression, they require measurements of cell shape and molecular counts. These may come from paired histology and sequencing or a combination of cell segmentation and count measurements from \emph{in situ} flouresence. In addition to our count matrix $X$ from earlier, let $\mathcal{S}$ containing $s_i \in {1, 2, \dots, N}$ cells represent the measurements of morphology, generally images or derived features. An experiment has captured a realization of the distribution over morphology, cell positions, and cell molecular states $p(\mathcal{S}, X, C).$



The analysis methods that currently exist make two strong assumptions: first, that each cell's expression and morphology are independent, and second that the position of the cell in space does not affect the morphology, in sum $p(s_i, x_i | s_{-1}, x_{-1}, C) = p(s_i, x_i)$. Then, one set of measurement is defined as a function of the other; shape as a function of gene expression, $p(s_i|x_i) = f(x_i)$, or gene expression as a function of shape, $p(x_i|s_i) = f(s_i)$. This is a reasonable assumption to make with current data and suggests a tractable class of model, but it obscures the complexity of the underlying mechanobiology that considers both internal cell state and environmental factors in cell morphology~\cite{yeung2005effects}. 

A natural opportunity in this space is to jointly model morphology and expression together, possibly by representing morphology using functional data analyses~\cite{meng2022randomness} or an autoencoder. Within a latent variable model, we may learn a shared representation of both cellular state and some encoding of cellular morphology $Z$ given some form $p(s_i, x_i|z_i) = f(z_i)$. Canonical correlation analysis, for example, has been used to jointly learn embeddings of gene expression and histology images for bulk RNA-seq data~\cite{ash2021joint}. Given sufficient single-cell data for network training, similar methods could be used to capture the relationship at single cell resolution without directionality assumptions. 

More intriguing are models that are able to capture the effect of nearby cell morphology and expression, similar to the signaling models explored before. A simple model would decompose the likelihood of expression and morphology $p(x_i, s_j| x_{-i}, s_{-i}, C)$, into terms representing the intrinsic cell morphology and deviations induced by environmental effects. With appropriate data, one could imagine more sophisticated models that are able to account for the organization of cells alongside their shapes and expression, a full joint model of $p(X, \mathcal{S}, C).$ Models of this type will likely require multiple replicates, both technical and biological, of spatial experiments to accurately estimate these distributions. But the increased use of spatial experiments and expanded fields-of-view in each sample will open these avenues for investigation. 

Returning to our second data representation---the list of molecular locations, identities, and cellular groupings---the opportunity exists to model the molecular levels effects on morphology and, conversely, the change in spatial distribution of molecules given morphology. A simple model might rely on the assumption of independence between cells, and posit that $p(s_i|m_i, c_i) = f(m_i, c_i).$ The correct functional form will depend on the representation of the shapes in $\mathcal{S}$; some tabular featurization can take advantage of regression models while a full image might require a neural network. Most exciting would be a model that can capture biophysical properties of the molecular interaction, learning how specific proteins or RNA molecules lead to the formation and warping of individual cell parts such as membrane structures within and across cells. Natural extensions would jointly consider niches of cells or full tissues, a model of $p(\mathcal{S}, M, O, C).$ 

For biologists who study dynamic processes such as development or cell response, time-dependent models will be the key to answering scientific questions. The desired model will include the evolution of expression and morphology as a function of time, $p(\mathcal{S}, X, C)(t)$ or $p(\mathcal{S}, M, O, C)(t)$. These models, coupled with appropriate data, may untangle the order in which morphology changes drive expression or expression changes morphology. Fitting models with clear biophysical interpretations may be one strategy to obtaining interpretable results, e.g. estimating the mechanical forces contributed from membrane proteins on maintaining rigidity. Combining flexible machine learning methods with a biophysical interpretation will likely be required to fully capture the complexity of these dynamic morphological processes.

\subsection*{III. Methods to investigate how cellular environment shapes cellular state, cellular division, cell differentiation, and clonal dynamics}

An exciting future direction is to map existing lineage-tracing methods onto spatial coordinates to better understand the spatial distribution and behavior of clones. Within our hypothetical framework, let us imagine that we have a count matrix $X$ and spatial coordinates $C$, as well as some additional data structure $Q$ that defines the relationship between cells (i.e., mother-daughter relationships in cells or cells that are part of one clonal population). One way to represent the ancestry of cells is by making $Q$ an adjacency matrix that represents a directed tree, where $Q_{ij} = 1$ if cell $i$ is a daughter of cell $j$. Connected components in this graph represent clonal outgrowth, and can be traced back to a single progenitor.

Although current analyses can identify clonal population sizes, it remains an open question whether these sizes are governed by cell-intrinsic or cell-extrinsic factors. If a set of cells $Y$ represent a connected component of $Q$, we can identify generations at which clonal expansion slowed or halted, and ask whether clonal size (the cardinality of $Y$, $|Y|$) is a function of expression in surrounding cells, $|Y| = f_{1}(x_{-Y}, c_{-Y}) + f_{2}(x_{Y}),$ where $x_{Y}, x_{-Y}$ are the expression profiles of cells in $Y$ and all cells not in $Y$, respectively. Similar to our spatial signaling framework, this treatment decouples the effects on clonal population size into clonal effects and the effects of environment around the clone.

Using this framework, we can also ask spatial questions about cells within a single clonal lineage: as described earlier, single-cell sequencing is able to resolve these populations but, before spatial sequencing, was unable to resolve their location. In some tissues, clones of cells remain close to each other in space and share a common niche. However, it is also possible for clones to split, migrate away from each other, and otherwise disperse in space. Given a set $Y$ of clones originating from a single cell, we can now study their dispersion patterns across space. Taking inspiration from our discussion of spatial signaling limits in cells, we can define a radius $r$ and compute the probability of a given clone lying within radius $r$ from other clones in the population:
$$p(dist(c_{i}, c_{j}) \leq r | i, j \in Y).$$
We can also ask whether this colocalization is more, less, or equally likely if cells come from the same clone. This value can be calculated and compared for multiple clonal populations $q_{1}, q_{2}, \dots$ to identify clone-specific spatial distributions and behaviors of daughter cells to stay close or intentionally disperse.
If there are members of a clonal lineage that are clearly separated in space, we can then ask how this stratification may have occurred as a function of cell state as well as the local cell population: $p(dist(c_{i}, c_{j}) \leq r | i, j \in Y) = f(x_{i, j}, x_{-i, -j}, c).$

With sufficient spatial genomic data, learning the function $f$ would most likely give higher weight to cells closer to the clones of interest, while also capturing environmental factors that define spatial clonal heterogeneity. The driving factors behind this spatial segregation may also be differentiation in the clones themselves; for instance, in the layers of the epidermis, cells from a single clone differentiate as they stratify from basal to apical~\cite{lechler2005asymmetric}. In this case, spatial segregation may largely be a function of the intrinsic cell state within clones $x_{i, j}$, and these effects, too, can be decoupled from effects from local cells.

\subsection*{IV. Methods to understand the relationship between cellular environment and rare events}

A number of methods are needed to resolve the relationship between cellular environment and rare events. First, identification of rare events is essential but challenging given current pipelines. Currently, rare events are often filtered or overlooked in spatial transcriptomic data. For example, jackpot cells likely will not be identified because of the large numbers of zeros in marker transcripts of rare cell types across all cells, leading to marker genes being removed from the analysis and preventing identification of rare cell types. The opportunity here is to work with the mapped but not filtered data to identify rare cell types through rare marker gene profiles. 

Second, understanding the environment characteristics that lead to rare events requires phenotyping a cellular environment and testing for enrichment of rare events within specific types of cellular environments. A number of methods perform adjacent analyses, quantifying differential cell-type adjacency across a tissue~\cite{cable2022cell}, functional cellular collectives~\cite{schurch2020coordinated,xu2022deepst,shi2023identifying}, and identifying de novo spatial domains~\cite{velten2022identifying,townes2023nonnegative}.

Third, identifying enrichment of specific rare events within a cellular environment may be challenging given the paucity of these rare events and the complexity of a cellular environment. Outlier detection methods may be useful in this space, but these methods are broad; in the context of probabilistic models, identifying cells that have a low probability of being generated from a foundational model or latent space model of diverse cells may suggest a rare cell type or cell state~\cite{theodoris2023transfer,verma2020robust,lopez2018deep}. A marked Poisson process may be useful to identify enrichment of specific environments in which these rare cell types arise. Marked Poisson processes consider specific events (here, a rare cell type) in the context of time or space with a ``mark'' or an identifier; then specific marks will filter up as enriched for rare events.

\section*{Concluding remarks}
The rapid development of spatial genomics technologies, for the first time combining spatial imaging of cells and tissues with an analysis of their state and genomic profiles,  provides an opportunity to revisit the types of questions we are able to ask and the quantitative methods we use to answer those questions. 


Here, we present four fundamental biological questions, each with profound implications for health and disease, that can now be addressed using spatial genomics technologies combined with appropriate machine learning methods. Future work will build on existing spatial genomics technologies and tailored analyses through the integration of time series data, better predictions of short-range and long-range correlations in multi-omic spatial datasets, and the ability to reason about biological processes across many scales.


\begin{backmatter}

\section*{Ethics approval and consent to participate}
No human subjects were involved in this paper.

\section*{Availability of data and material}
No new data and materials were produced for this paper. Referenced papers are available through their publishers.

\section*{Competing interests}
BEE is on the SAB of Creyon Bio, Arrepath, and Freenome; BEE consults for Neumora. AV consults for NE47 Bio. 

\section*{Funding}
BEE and AV were funded by Helmsley Trust grant AWD1006624, NIH NCI 5U2CCA233195, CZI and CIFAR under the Multiscale Human Program.

\section*{Author's contributions}
SGJ, AV, and BEE drafted, wrote, and edited this manuscript.

\section*{Acknowledgements}
None


\bibliographystyle{bmc-mathphys} 
\bibliography{bmc_article}      


\end{backmatter}
\end{document}